# Wi-Fi Sensing: Applications and Challenges

**A. M. Khalili [1,*], Abdel-Hamid Soliman [1], Md Asaduzzaman [1], Alison Griffiths [1]**

[1] School of Creative Arts and Engineering, Staffordshire University, United Kingdom;
* Correspondence: a.m.khalili@outlook.com

**Abstract:** Wi-Fi technology has strong potentials in indoor and outdoor sensing applications, it has several important features which makes it an appealing option compared to other sensing technologies. This paper presents a survey on different applications of Wi-Fi based sensing systems such as elderly people monitoring, activity classification, gesture recognition, people counting, through the wall sensing, behind the corner sensing, and many other applications. The challenges and interesting future directions are also highlighted.

---

## 1. Introduction

The discovery of the wireless wave by Hertz [1] has opened the doors for many technological revolutions. Most aspects of our modern life have been affected by this important discovery. Today, 130 years after Hertz's discovery, there is a wide range of applications of the wireless waves, such as activity detection, gesture recognition, and elderly people monitoring. Future access points will be also able to recognise gestures and take commands, analyse and classify different activities of people inside and outside the house, monitor the health conditions of elderly people by monitoring their breath, fall, etc. Table .1 and Fig. 1 summarises the applications of using the Wi-Fi signals. The range of applications is not limited to indoor applications but also include outdoor areas.

**Table 1.** Different applications of Wi-Fi based sensing systems.

| **Applications** | **References** |
|---|---|
| Health Monitoring | [55-56], [169-177]. |
| Activity Classification | [68-76], [178]. |
| Gesture Recognition | [82-86], [179-181]. |
| People Counting | [87-102], [160-161]. |
| Through the Wall Sensing | [109-112], [182-183]. |
| Emotion Recognition | [139]. |
| Attention Monitoring | [147]. |
| Keystrokes Recognition | [153]. |
| Drawing in the Air | [154], [184]. |
| Imaging | [155], [185-186]. |
| Step Counting | [156, 157], [187]. |

| Speed Estimation | [158]. |
|---|---|
| Sleep Detection | [188, 190]. |
| Traffic Monitoring | [191]. |
| Smoking Detection | [192-193]. |
| Metal Detection | [194-195]. |
| Sign Language Recognition | [196-199]. |
| Humidity Estimation | [200]. |
| Wheat Moisture Detection | [201]. |
| Fruit Ripeness Detection | [202]. |

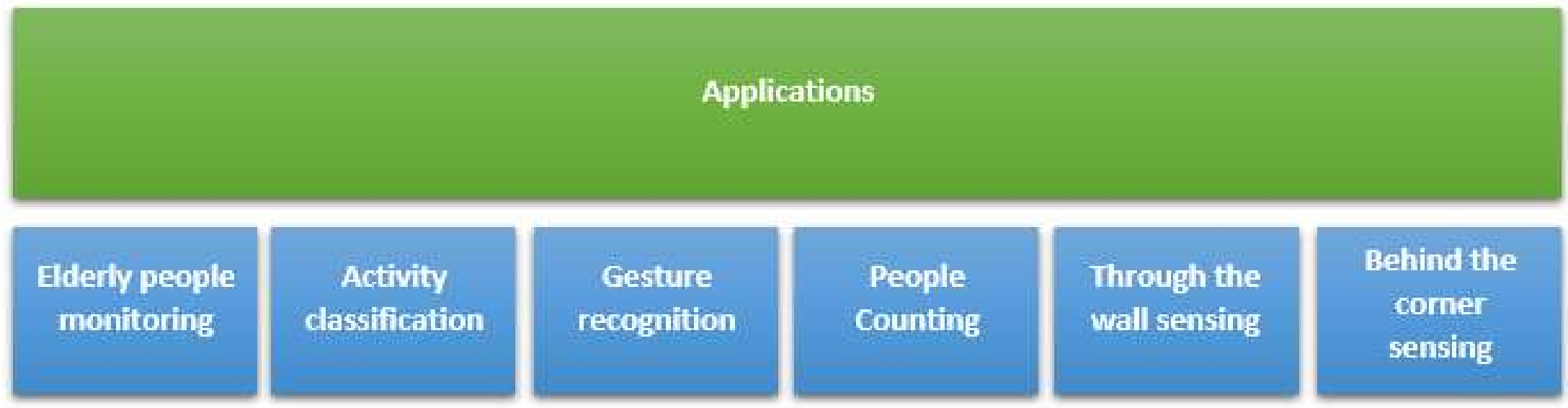

**Fig 1.** Different applications of Wi-Fi based sensing systems.

Video-based sensing systems have limitations in Non-Line-Of-Sight (NLOS) environments, in the dark, through smoke or walls; they are also computationally intensive and have lower localisation accuracy.

Wi-Fi does not have the limitations of video-based systems; furthermore, it has higher availability and longer range than other signal-based systems such as Ultra-Wideband (UWB). The possibility to provide people tracking without transmitting any additional signal, nor require co-operative objects as other signal-based systems, offers major opportunities.

In this paper, we extend and build on recent works such as [9-11] to provide a more comprehensive survey on recent applications of Wi-Fi based sensing systems. The main used approaches of Wi-Fi based sensing systems with their limitations are summarised. Then, a survey on different recent applications is presented.

The paper is organized as follows: The main used approaches of Wi-Fi based sensing systems with their limitations are described in section 2. The application of the Wi-Fi signal in elderly people monitoring is described in section 3.1. The application of the Wi-Fi signal in activity classification is described in section 3.2. Section 3.3 describes the application of the Wi-Fi signal in gesture recognition. The application of the Wi-Fi signal in people counting is described in section 3.4. The application of the Wi-Fi signal in through the wall sensing is described in section 3.5. The application of the Wi-Fi signal in behind the corner sensing is described in section 3.6. Other Applications are described in section 3.7. The gaps, limitations, and interesting future directions are described in section 4, and the paper is concluded in section 5.

## 2. Wi-Fi based sensing systems

Vision-based people tracking systems [12, 13] have been widely used recently for different applications such as activity classification, gesture recognition, elderly people monitoring, and people counting. However, these systems have many limitations in Non-Line-Of-Sight (NLOS) environments, in dark, through smoke or walls; they are also computationally intensive and have lower localisation accuracy.

Traditional radar systems have been recently used to perform people tracking and activity recognition [14, 15]. However, these systems use multiple antennas, and expensive ultra-wideband transceivers.

Within the European project ATOM [16, 17, 18], the potential of Wi-Fi for people tracking within different public areas such as airport terminals was investigated. The use of the Wi-Fi signals turned out to be very promising, where Wi-Fi signals represent a very suitable solution for the following reasons:

- Reasonable bandwidth, which will lead to a high range resolution.

- Wide coverage, Wi-Fi networks are spreading at a very high rate for both commercial and private use.

- Reasonable Transmitted power, which gives the Wi-Fi signal an advantage over short-range sensing technology such as UWB.

Colone et al. [17] investigated the use of Wi-Fi signals for people tracking, they conducted an ambiguity function analysis for Wi-Fi signals. They also investigated the range resolution for both the Direct Sequence Spread Spectrum (DSSS) and the Orthogonal Frequency Division Multiplexing (OFDM) frames, for both the range and the Doppler dimensions, large sidelobes were detected, which explains the masking of closely spaced users. Falcone et al. [18] presented the results of detecting the speed and the range of a car by correlating the received Wi-Fi signal with the transmitted one. It was shown that the moving car can be localised, but the user next to it is masked by the strong reflection from the car. Then they showed that ambiguity function control filter and disturbance removal techniques could allow the detection of both the car and the person.

Wi-Fi based systems use the variations in the wireless channel to track people in a given environment. Existing systems can be grouped into three main categories: (1) Received Signal Strength (RSS) based, (2) Channel State Information (CSI) based, and (3) Software Defined Radio (SDR) based.

RSS provides only coarse-grained information about the variations of the wireless channel and does not provide fine-grained information about the multipath effects. CSI was introduced to capture fine-grained variations in the wireless channel. Received signal strength measurements are only a single value per packet, which represents Signal-to-Interference-Noise Ratio (SINR) over the channel, channel state information on the other hand contains the amplitude and the phase measurements for each OFDM subcarrier. SDR based systems are low-level systems that have full access to the received signal and therefore can capture more valuable information from the received signal.

In the SDR based category, the first experiments to localise people using Wi-Fi signals were done by Guo et al. [19]. The Wi-Fi signals were utilised for localisation by matching the transmitted signal with the received one, the localisation of one person was achieved in an open field without much clutter. Chetty et al. [20] conducted experiments in high clutter indoor environments using Wi-Fi signals, they were able to detect one moving person through a wall.

A multi-person localisation system Wi-Track [21] was proposed by Adib et al. They pinpoint users' locations based on the reflections of Wi-Fi signals off the persons' bodies, their results show that their system can localise up to five users at the same time with an average accuracy of 11.7 cm. The system uses the reflection of the signal to estimate the time required by the signal to travel from the antennas to the person and back to the antennas. The system then uses the information of the antennas' positions to build a geometric model that converts the round trip delays of the reflected signal to a position of the user. Wi-Track removes the reflections from walls and other static objects by background subtraction where the distance of these objects does not vary over time, and hence they can be removed by subtracting consecutive frames of the constructed scenes. Reflections that include a combination of humans and static objects are addressed through taking into account the models of human's motion and their velocity in indoor scenarios. One limitation of the proposed system is that it needs the users to move in order to be able to locate them because the system cannot distinguish between static users and a piece of furniture.

When a person is conducting an activity he will cause the blocking or the reflecting of transmitted signals. This will cause a variation in the received signal strength. The activities performed by people will leave a characteristic fingerprint on the received signals. The variation in the received signal can then be used in order to classify different activities. Woyach et al. [22] investigated the effect of human's motion on the received signal. Moreover, they showed that the speed of an object could be estimated by analysing the pattern of RSS variation of transmitted frames of a moving object. Krishnan et al. [23] expanded the work of Woyach by studying the differences between moving objects and stationary objects by analysing the variation of the RSS in a network of wireless nodes. Anderson et al. [24] and Sohn et al. [25] were able to distinguish between six speed levels.

Youssef et al. [26] introduced a Device-free Passive (DFP) localisation system. A DFP system can localise objects that do not carry any device. The system works by observing variations in the received signals to detect the presence of objects in the environment. Bocca et al. [27] proposed a DFP system which localises the person based on the RSS variations of a line of sight link between two communication nodes, a sub-meter accuracy was reported; however, these methods have serious limitations in non-LOS environments due to the multipath effects. For non-LOS environments, Wilson et al. [28] also proposed a variance-based method to localise people; however, their method cannot locate static people, since they do not produce much RSS variance.

A more recent work by Wilson et al. [29] investigated the use of the particle filter to localise both static and moving people. The method works in both LOS and non-LOS environments; however, it cannot be easily implemented in real-time. Furthermore, the accuracy of RSS based methods requires a high density of communication nodes.

Kosba et al. [30] proposed a system to detect motion using standard Wi-Fi hardware. Their system uses an offline training phase where no movement is assumed as a baseline. Then, the anomaly is detected by detecting changes from the baseline. Lee et al. [31] also used the RSS fluctuation of communication nodes for intrusion detection. They reported changes in the standard deviation and the mean of RSS values in five distinct indoor scenarios.

RSS is an unreliable measure, because it is roughly measured, and can be easily affected by multipath. In [32] the Channel State Information is used, CSI is a fine-grained information, it gives information about the frequency diversity characteristic of the OFDM systems. In [32] the authors used the CSI to build an indoor localisation system FILA. FILA processes the CSI of multiple subcarriers in one packet and builds a propagation model that captures the relation between CSI and the distance. The effectiveness of the system is shown by using a commercial 802.11n device. Then, a series of experiments were conducted to evaluate the performance of the proposed system in indoor environments. The experiments results showed that the localisation accuracy could be significantly improved by using CSI, where for over 90% of the data points, the localisation error was in the range of 1 meter.

Authors in [33] showed that activity recognition can be achieved using the CSI measurements which are available by the IEEE 802.11n devices and with a small number of communication nodes. Their system E-eyes uses the wide bandwidth of 802.11ac, where a more fine-grained channel state information is used in Multiple Input Multiple Output (MIMO) communications. Different sub-carriers will encounter different multipath fading because of the small frequency difference. When taking a single RSS measurement, such effect is usually averaged out. Each subcarrier measurement will change when a movement changes the multipath environment. This will allow the system not only to detect changes in the direct path but also to take advantage of the rich reflected signals to cover the space. This will also allow the system to operate using one access point and a small number of Wi-Fi devices, which already exist in many buildings. However, the proposed system has many limitations: first, the system was designed and tested with the presence of only one person. Second, the system requires a stable surrounding environment with no furniture movement, because changing the surrounding environment requires a profile update.

## 3. Applications of Wi-Fi based people tracking systems

### 3.1 Elderly people monitoring

The population of people aged 65 years or older is increasing, and their ratio to the population of people aged 20–64 will approach 35% in 2030 [34]. Every year 33% of elderly people over the age of 65 will fall [34]. The fall could cause injuries and reduction of the quality of life, fall represents one of the main reason of the death of elderly people.

Authors in [55] proposed a Wi-Fi based fall detection system WiFall, by taking advantage of the channel state information measurements. The basic idea is to analyse the change in CSI when human activities affect the environment. The system consists of two stages: the first one is an algorithm to detect abnormal CSI series, and the second one is an activity classification based on Support Vector

Machine (SVM) technique to distinguish falls from other activities. WiFall achieved comparable precision to device-based fall detection systems with 87% detection rate and 18% false alarm rate.

Patwari et al. [56] reported that they were able to detect the breathing rate of a person by analysing the fluctuation of RSS in the received packets from 20 nodes around the person. By using the maximum likelihood estimation, the breathing rate was estimated with an error of 0.3 breaths per minute. The nodes transmit every 240ms with a 2.48 GHz frequency, which means that the overall transmission rate is about 4.16Hz. The prediction was performed after a 10 to 60 second measurement period. Longer measurement periods did not significantly improve the accuracy. The achieved accuracy was related to the number of nodes, where with 7 nodes, an RMSE rate of 1.0 approximately was achieved.

**3.2 Activity classification**

The growing concern about law enforcement and public safety has resulted in a large increase in the number of surveillance cameras. There is a growing interest in both the research community and in the industry to automate the analysis of human activities and behaviours. The main approach of these techniques is to model normal behaviours, and then detecting the abnormal behaviour by comparing the observed behaviour and the normal behaviour. Then the variation is labelled as abnormal. Abnormal behaviour detection has gained increasing interest in surveillance applications recently. Hu [57] has recently discussed that most surveillance techniques are based on the same approach; where the moving object is first detected. After that, it is tracked over many frames, and finally the resulted path is used to differentiate normal behaviour from abnormal ones. In general, these techniques have a training stage where a probabilistic model is built based on the normal behaviour.

Authors of [68] classified simple activities by capturing features from the variation of the signal between two communication nodes. They also investigated the performance of the system under multipath environments. It was also demonstrated that activities conducted at the same time from multiple persons could be easily distinguished by using signal strength based features [69]. However, the highest classification accuracy was achieved when the activity was less than one meters from the receiver. At larger distances, the classification accuracy decreased rapidly. Recently in [70], they considered the recognition of general activities based on RSS in a sensor network, where activities such as sitting, standing, walking, and lying have been recognised with high accuracy.

Authors in [71] proposed the use of the wireless channel, where they monitored the fluctuation in the RSS, which is calculated for each packet at the receiver, they attempted to recognise activities performed in front of a mobile phone. This approach allows activity recognition when the device is not carried by the person but near to him. The achieved accuracy is still below the accuracy of conventional sensors such as accelerometers, where an accuracy of 74% inside the room and an accuracy of 61% through the wall were achieved.

For activity recognition, many simple RSS features have been used such as average magnitude squared, signal-to-noise ratio [72], [73], [74], and signal amplitude [75]. In [76] the learning approach was able to detect and count up to 10 moving or stationary users. Then, after using additional frequency domain features, the accuracy was further improved [77]. In [78] the authors proposed a system that can recognise the gesture of multiple users.

**3.3 Gesture recognition**

As computers become increasingly embedded in the environments, there is an increasing need for novel ways to interact with the computers [84]. For instance, by a hand motion in the air, the person can adjust the volume of the music. Such capabilities can enable applications in many domains including gaming, home automation, and elderly health care as described in Fig 2. Conventional gesture recognition systems are based either on vision technology such as Kinect or wearable sensors such as Magicrings.

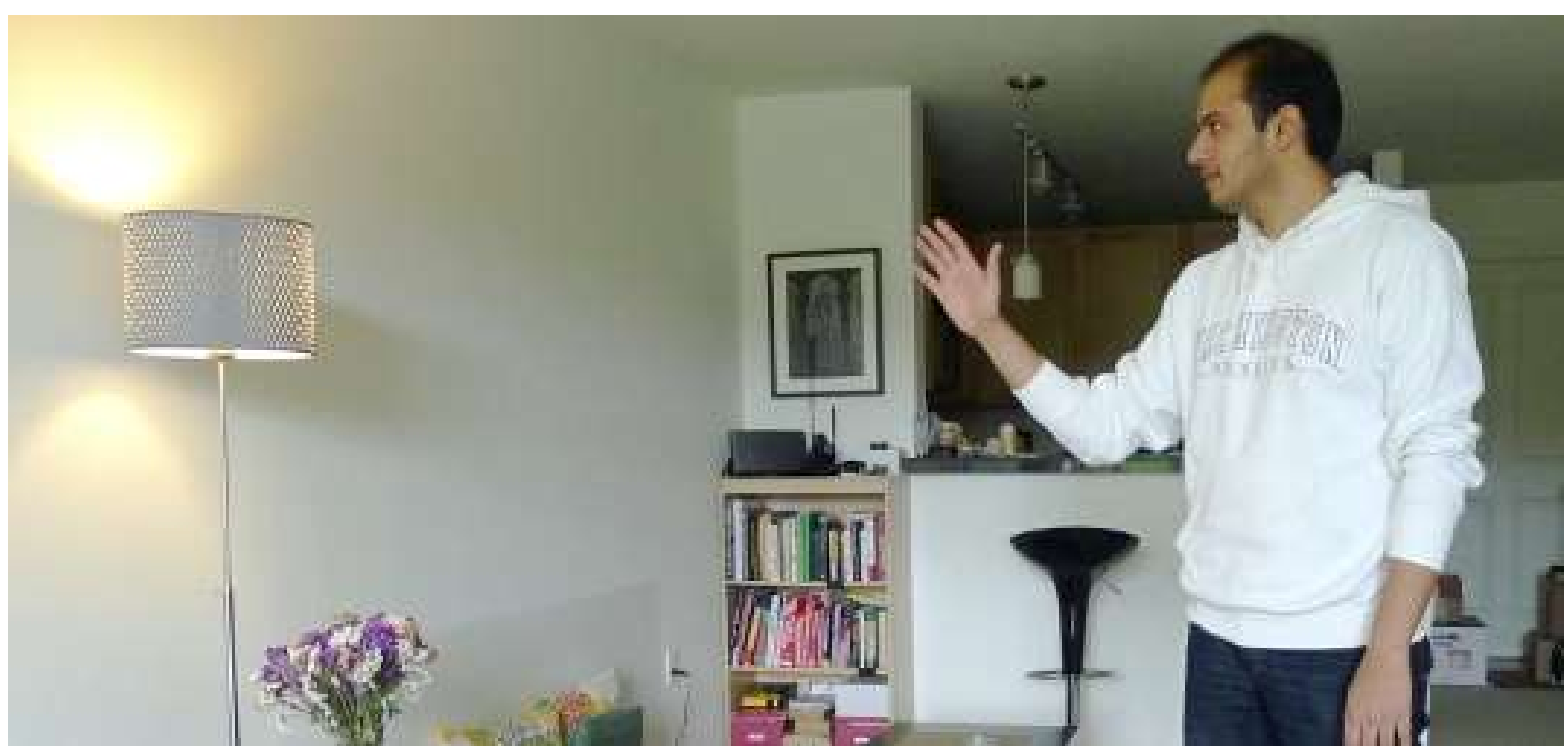

Fig 2. Application of gesture recognition in home automation [84].

Aumi et al [80] presented an ultrasonic-based gesture recognition approach. It uses the integrated audio hardware in smartphones to determine if a particular phone is being pointed at, i.e., the person waves at a phone in a pointing motion. They evaluated the accuracy of the system in a controlled environment. The results show that, within 3 meters, the system has an accuracy of 95% for device selection. The basic idea of the proposed system is that the intended target phone will have the maximum Doppler shift compared to the other potential target phones. By comparing the peak Doppler shift in all the phones, they can determine the intended phone.

Gupta et al [81] presented SoundWave, a gestures recognition system that uses the microphone and the speaker that are already integrated into most smartphones to recognise gestures around the phone. They generated an inaudible tone, which will have Doppler shift when it bounces off moving objects such as the hand. They calculated this Doppler shift using the microphone to recognise different gestures.

Abdelnasser et al. [82] presented a Wi-Fi based gesture recognition system by using variation in RSS resulting from hand gestures. The system can recognise many hand gestures and translate them into commands to control different applications. The gesture recognition accuracy was 87.5% when a single access point was used and 96% when three access points were used. However, RSS is not an accurate metric because the high variation in RSS measurements causes a high rate of misdetection.

Cohn et al. [83] used the electromagnetic noise resulted from electronic devices to recognise different gestures. They presented accurate gesture recognition with an accuracy of 93% for 12 gestures. They also presented promising results for people localisation inside a building. They used variations in the received signal that happen when the body moves. In addition to the ability to recognise different whole-body gestures, they also showed accurate localisation of the person within the building based on a set of trained locations. Their system was based on electromagnetic noise resulted from electronic devices and the power lines. However, the system requires the user to train and calibrate the gestures and locations for his home, the classification works well if the home is in the same state during the training; however, large changes in the state (such as turning on the lights) drop the classification accuracy significantly. Some devices also generate broadband noise that might mask other noise signals.

Gupta et al. [84] proposed WISEE, a gesture recognition system that uses Wi-Fi signals to recognise human gesture. WISEE can recognise user gestures without introducing any additional sensing device on the user body. The system uses the Doppler shift, which is the frequency change of the wireless wave when its source moves toward the observer. There will be many reflections from the user body, and the user gestures will result in a certain pattern of Doppler shift. For instance, if the user moves away from the device, this will produce a negative Doppler shift, and if the user moves toward the device, this will produce a positive Doppler shift as described by Fig 3.

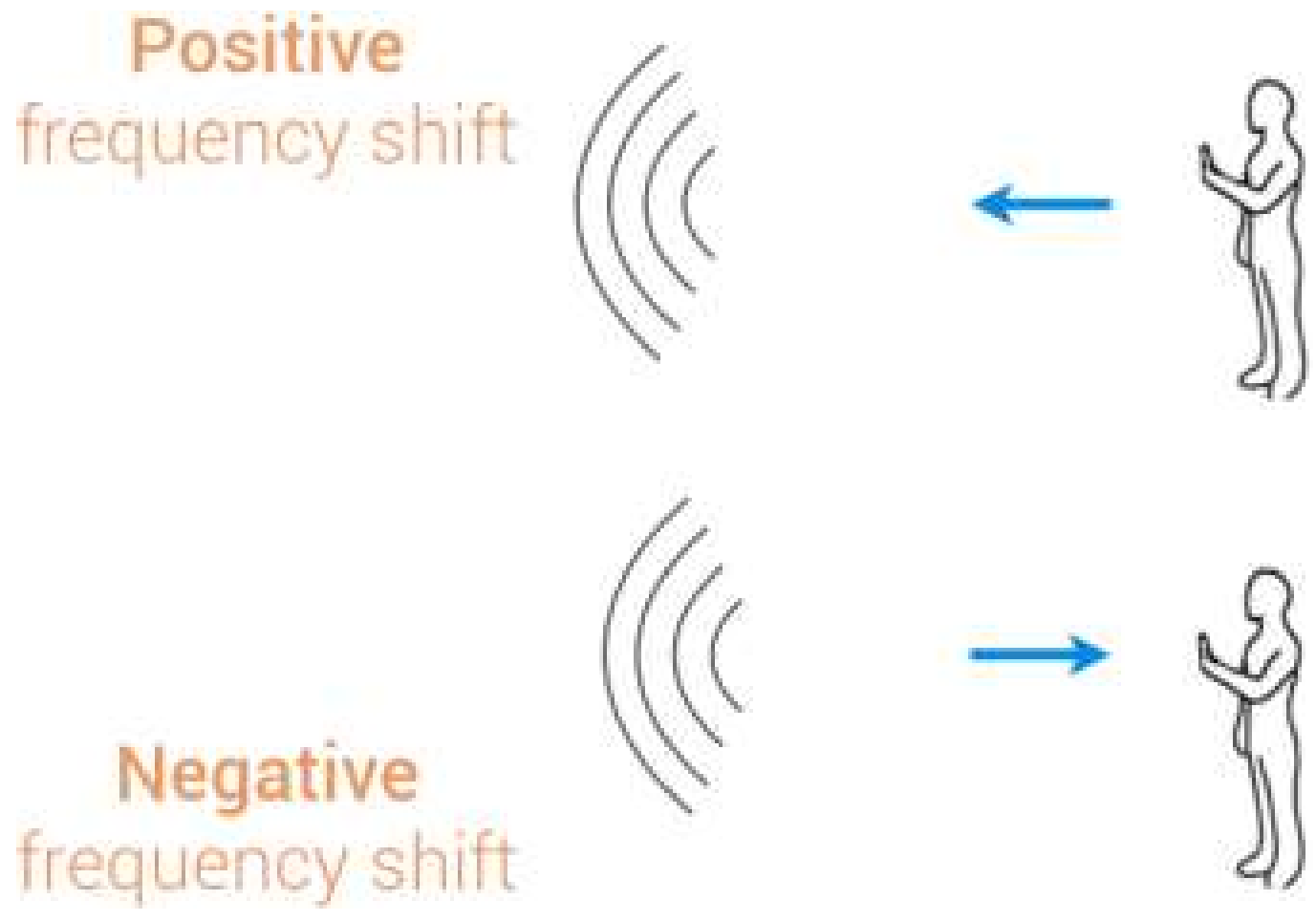

Fig 3. The effect of the movement direction on the Doppler shift [84].

The main challenge for the proposed system was that the user gesture produces very tiny changes in Doppler shifts, which is very difficult to detect using WI-FI signals. A movement of 0.5 m/sec produces 17 Hz Doppler shift if the 5 GHz frequency is used. For gesture recognition applications, a Doppler shift of few Hertz should be detected. The solution for this challenge was achieved by converting the received signal, which is reflected from the moving object to a narrowband signal with few Hertz bandwidth, then the system extracts the frequency of this signal to recognise small Doppler shifts. The results of classifying 9 gestures in LOS and NLOS environments show that 94% of gestures were classified correctly and 2% of gestures were not detected.

Wang et al. [85] presented WiHear, which investigated the potential of using Wi-Fi signals to hear the talk of people. The proposed system locates the mouth of the user and then recognises his speech by analysing the signals reflected from his mouth. By analysing the mouth moving patterns, the system can recognise words in a similar way to lips reading. The results show that using a pre-defined vocabulary, the system can achieve recognition accuracy of 91% for single user speaking no more than 6 words and 74% accuracy for no more than 3 people speaking at the same time. The accuracy decreases when the number of persons increases. Furthermore, the accuracy decreases dramatically when more than 6 words were spoken by each user. The system also assumes that people do not move while they are speaking, and the recognition accuracy of 18% is very low for through the wall scenarios. In [86] the authors proposed a system that can recognise the gesture of multiple users.

### 3.4 People counting

Crowd counting is increasingly becoming important in a number of applications, such as crowd control and guided tour [87]. There are many applications that can benefit from people counting. Smart building management is one example, where the heating can be optimised based on the number of people, which can result in a large energy saving.

Mostofi et al. [87] proposed a Wi-Fi based system that counts the number of walking people in an area using only RSS measurements between a pair of transmitter and receiver antennas. The proposed framework is based on two important ways that people affect the propagation of the Wi-Fi signal, the first one is by blocking the line of sight signal, and the second one is the scattering effects. They developed a basic motion model, then they described mathematically the effect of a crowd on blocking the line of sight. Finally, they described mathematically the effect of the number of people on the resulted multipath fading and the scattering effects. By integrating these two effects together, they were able to develop a mathematical equation describing the probability distribution of the received signal amplitude in term of the number of people. In order to test the proposed approach, large outdoor and indoor experiments were conducted to count up to 9 persons as described by Fig 4, the results show that the proposed approach can count the number of persons with a high accuracy using only one Wi-Fi transmitter and one Wi-Fi receiver. For example, an error of 2 or less was achieved 63% of the time for the indoor case, and 96% of the time for the outdoor case when using the standard Wi-Fi omnidirectional antennas. When directional antennas were used, an error of 2 or less was achieved 100% of the time for both the indoor and outdoor cases.

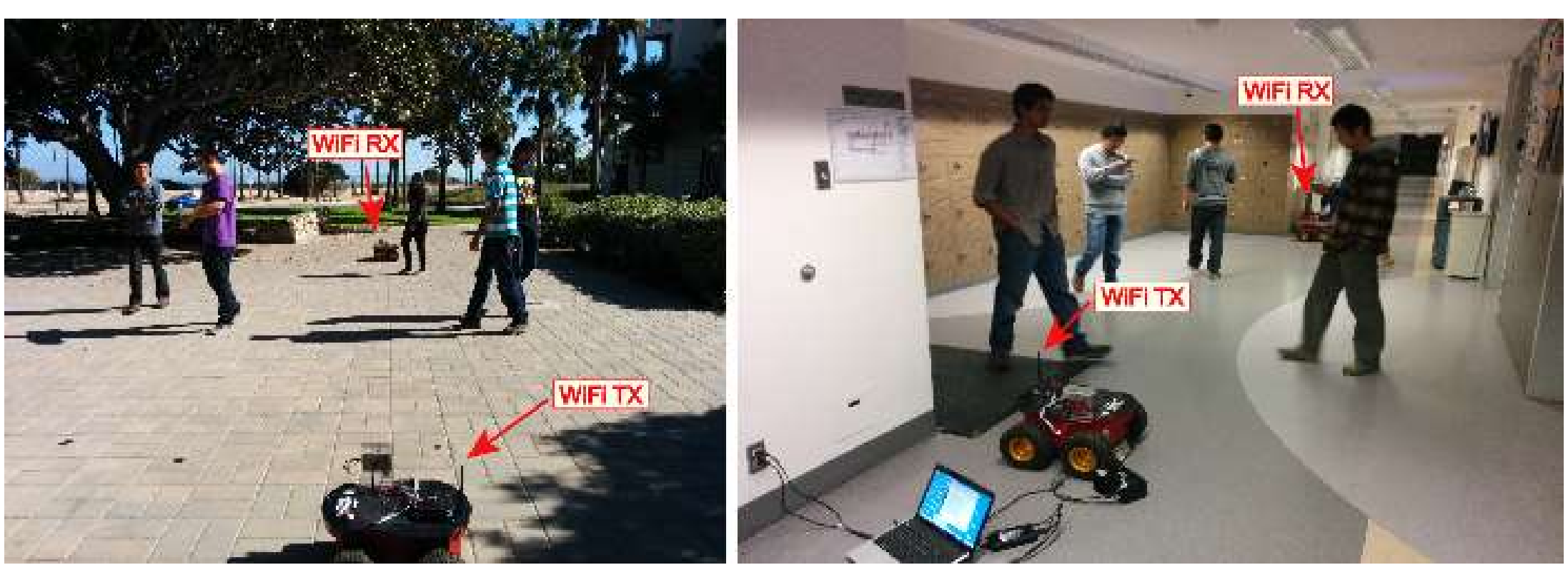

Fig 4. People counting in outdoor and indoor environments using only one pair of Wi-Fi cards [79].

In [88], multiple Wi-Fi nodes and RSS measurements were used to count the number of up to 4 persons. They reported an accuracy within an error of 1 person 84% of the time approximately. In [89] a similar approach was used but with fewer nodes, they were able to count up to three people. In [90], a transmitter-receiver pair was used to estimate the number of people based on RSS measurements. An extensive training data was used to develop the underlying model, an error up to 6 persons were reported in experiments limited to 9 persons.

In [91], the authors measured the channel state information of different sub-carriers, they developed a model to relate the channel state information to the number of persons through a training stage. They tested their model using one transmitter and three receivers to count up to 9 persons. However, measuring channel state information of different sub-bands is not available for most current Wi-Fi cards. In [92], the authors used UWB radar to count up to 3 stationary persons behind walls. In [93], the authors used a pulsed radar to estimate the number of people by using machine learning techniques.

Xi et al. [94] proposed a people counting system based on channel state information measurements. The basic idea of the proposed approach is that the number of people can be accurately estimated by analysing the changes in the channel state information. They theoretically studied and experimentally validated the relationship between the variation of the wireless channel and the number of moving persons. Their results show that CSI is very sensitive to the influence of the environment, they also showed that there is a monotonic relation between the number of moving persons and CSI variations. This provides a solid ground for crowd counting. They proposed a metric, which is the percentage of non-zero elements in the CSI Matrix. To estimate the number of people, the metric can measure the changes in CSI in a very short time. The value of the metric increases as the number of active persons increases, and it reaches the saturated state when the number of persons reaches a certain threshold. The Grey-Verhulst model was applied to estimate the number of persons. To estimate the number of persons in a large area, multiple devices were used to form a grid array. The main challenge was that CSI is very sensitive to the environment, i.e., users moving in one grid will result in CSI variations in adjacent grids. To address this challenge, an interference cancelation technique was proposed to adjust the sensing range for each receiver to enhance the estimation accuracy in a large monitored area. The system was built using 802.11n Wi-Fi devices. The system was evaluated with large-scale experiments. The results showed that the proposed approach outperforms other approaches in terms of accuracy and scalability.

In [95], [96] the locating procedure was divided into two stages: the training stage and the operating stage. Xu et al. [96] formulated the localisation problem as a probabilistic classification problem to cope with the error caused by the multipath in cluttered environments. Yuan et al. [95] used a classification algorithm to estimate the number of persons. Arai et al. [97] proposed an approach to link the crowd movement patterns with the feature of the radar chart. This approach requires a survey over the used areas to build a fingerprint database. The efforts, cost, inflexibility, and the environment dynamics are the main limitations of this approach. In crowd counting, the training cost is a main limiting factor particularly for large-scale scenarios; furthermore, it is very challenging to get the ground truth when the number of persons is large.

In [98], [99] if the person is nearby a link, the RSS will change remarkably. However, if the person moves away from the link, the performance decreases rapidly. Nakatsuka et al. [100] demonstrated the effectiveness of using the average and the variance of RSS to estimate the crowd density. Patwari et al. [101] proposed a statistical approach to model the RSS variance as a function of a person's position with respect to the antennas locations. Xu et al. [102] used a link-based approach to estimate the number of persons and locate their positions using RSS measurements.

### 3.5 Through the wall sensing

Through the wall sensing [103] is a new research area that was introduced to address the increasing need to see through the walls for many applications, such as recognising and classifying objects in the building. It could be also used in emergency situations such as earthquakes to check whether a person exists under the rubble. Through the wall sensing is highly desirable by emergency workers and the police, it can also help firefighters to locate people who are trapped inside a burning building.

Through the wall imaging has attracted much interest recently particularly for security applications [103]. Through the wall sensing systems must take into account signal attenuation caused by the walls, where the attenuation is lower at low frequencies. It must also take into account the need for large bandwidths to get a high range resolution. The majority of through the wall sensors are UWB radars, which have many advantages over classical narrow band sensors.

In [109] and [110] a series of experiments were conducted to investigate the effectiveness of using Wi-Fi signals as an illuminator of opportunity for through the wall people localisation. In [110] an indoor events detection system was proposed by using the time reversal technique to detect changes in indoor multipath environments. The proposed system enables a single antenna device that operates in the Industrial Scientific and Medical (ISM) band to capture indoor activities through the walls. The system uses the time reversal technique to detect changes in the environment and to compress high-dimensional features by mapping the multipath profile to the time reversal space, which will enable the implementation of fast and simple detection algorithms. Furthermore, a real prototype was built to evaluate the feasibility and the performance of the system. The experimental results showed that the system achieved a detection rate of 96.92% with a false alarm rate of less than 3.08% in both LOS and NLOS environments. However, when the person is close to the transmitter or the receiver, the miss detection rate increased significantly.

In [111] a new method for localisation and motion tracking through walls was presented. The method takes advantage of variations in received signal strength measurements caused by people motions. By using a model for the multipath channel, they showed that the signal strength of a wireless link is highly dependent on the multipath components that contain moving objects. A mathematical model relating the locations of movement to the RSS variance was used to estimate the motion. From that motion, the Kalman filter is then used to track the positions of the moving objects. The experimental results were presented for 34 nodes that perform through the wall tracking over an area that covers a 780 square foot. The system was able to track a moving object through the walls with a 3ft average error approximately. An object that moves in place can be localised with 1.5ft average error approximately.

Authors in [112] designed and implemented a through the wall people localisation system. Their methodology depends on detecting when people cross the links between the receivers and the transmitters. When two Wi-Fi 802.11n nodes were used, the methods achieved approximately 100% accuracy in detecting line crossings and movement direction. They also found that the proposed method achieved 90−100% accuracy when a single 802.11n receiver is used. However, the systems proposed in [111] and [112] require a large number of communication nodes which limits the range of applications of these systems.

**3.6 Behind the corner sensing**

Detecting and localising people situated behind obstacles could have many applications, obstacles might partially or completely block the propagation of wireless signals. Such situations may arise when for instance police forces want to inspect a corridor for possible threats before entering it. Wi-Fi has the potential for sensing behind corners using the diffraction and reflection of electromagnetic waves, for both indoors and outdoors applications.

Darpa developed a multipath exploitation radar program [113–115], the system tracks moving objects by utilising the multipath effect to maintain the track even when the objects are not in the line of sight. The same approach was used in [116–118] for behind the corner localisation of mobile terminals in urban environments. The multipath represented by the multiple echoes that are diffracted and reflected by an object and its surrounding environment are usually nuisance signals for conventional localisation systems [119, 120]. In [121], rather than considering them a nuisance, the multipath is used for localisation of people invisible to police forces. In addition to the reflection-based multipath, they used the diffraction and the combination of diffraction and reflection for the localisation. The proposed approach does not need a priori information about the geometry of the environment. It only needs information about the distance between the walls and the antenna, as well as the distance between the corner that diffracts the electromagnetic waves and the antenna. This information could be either obtained directly from the UWB data or extracted from other additional measurement devices. They showed results of successfully detecting and localising a person standing up to five meters away from the corner. The precision of the proposed approach depends on the size of the object. When only one single path is available, the localisation accuracy significantly decreases. One other limitation of the proposed approach is that the antenna should be directed toward the diffracting corner to increase the power of the diffracted path since it is very weak.

In [122], the authors showed that micro-Doppler signatures from person gait could be captured in an urban environment by using multipath propagation to illuminate the object in NLOS regions. A high-resolution radar system was used for data collection. The high resolution will enable multipath contributions to be separated individually. Two scenarios with 1 and 2 walking users are tested, the experimental was arranged to detect the multipath object response from up to 5 wall reflections. The main results showed that human micro-Doppler signatures could be used for classification purposes even after multiple wall reflections.

In [123] the authors have demonstrated the feasibility of X-band radar to detect moving persons behind concrete walls. The detection was achieved using stepped-frequency radar in a controlled scenario. Different measurements of the transmission and reflection properties of the material of the wall have indicated low transmission through the used wall type, leaving the diffracted, and reflected wave components as the main way for the interaction with objects behind the wall. However, the main challenge facing the proposed system is the multipath propagation.

**3.7 Other applications**

Emotion recognition is an active research area that has drawn growing interest recently from the research community [139]. Bull [133] indicated that dynamic configurations of the human body hold a large amount of information about emotions. He showed that body motions and positions could be used as an indicator of the human state such as boredom or interest along with other 14 emotions. Wi-Fi could play an interesting role to detect body pose and gesture, and to use this information to recognise human emotions.

The researchers in [139] presented a new system that can recognise user emotions using RF signals that are reflected off his body. The system transmits a wireless signal and analyses the reflections from the user body to recognise his emotions such as happiness, sadness, etc. The key building block of the system is a new algorithm that extracts the heartbeats from the wireless signal at an accuracy close to Electrocardiogram (ECG) monitors. The extracted heartbeats are then used to extract features related to emotions, then these features are used in a machine learning emotion classifier. The researchers demonstrated that the emotion recognition accuracy is comparable with the state of the art emotion recognition systems based on ECG monitors. The accuracy of emotion classification is 87% in the proposed system and 88.2% in the ECG based systems.

In [153], it was shown that Wi-Fi signals could be used to recognise keystrokes. Wi-Fi signals are now everywhere, at offices, home, and shopping centres. The basic idea is that while typing a specific key, the fingers and hands of the person move in a unique formation, and therefore produce a unique time-series pattern of channel state information values, which can be called the CSI waveform of that key. The keystrokes of the keys produce relatively different multipath variations in Wi-Fi signals, which can be used to recognise keystrokes. Due to the high data rates of recent Wi-Fi devices, Wi-Fi devices produce enough CSI values within the duration of a keystroke, which will help in building more accurate keystrokes recognition systems. In [153], a keystroke database of 10 human subjects was built. The keystroke detection rate of the proposed system was 97.5% and the recognition accuracy for classifying a single key was 96.4%. The proposed system can recognise keystrokes in a continuously typing situation with an accuracy of 93.5%. However, the system works well only in controlled environments. The accuracy of the system is affected by many factors such as changes in distance and orientation of transceivers, human motions in surrounding areas, typing speed, and keyboard size and layout.

In [154], it was demonstrated that it is possible to use Wi-Fi signals to enable hands-free drawing in the air. They introduced WiDraw, a hand tracking system that uses Wi-Fi signals to track the positions of the user's hand in both LOS and NLOS environments, without requiring the user to hold any device. The prototype used a wireless card, less than 5 cm error on average was reported in tracking the user's hand. They also used the same system to develop an in-air handwriting app, a word recognition accuracy of 91% was reported. However, one limitation of the proposed system is that it requires at least a dozen transmitters in order to be able to track the hand with high accuracy. Furthermore, the 3D tracking error is higher than the 2D tracking error, the main cause for this is the difficulty in accurately tracking depth changes. The system achieved high tracking accuracy only when the hand is within two feet from the receiver. The error starts to increase at larger distances.

The advantages and limits of performing imaging based on Wi-Fi signals were investigated in [155]. They presented Wision, a system that enables imaging of objects using Wi-Fi signals. The system uses the Wi-Fi signals from the environment to enable imaging. The approach uses multipath propagation where the signals reflect from objects before they arrive at the system. However, the main challenge is that the system receives a combination of reflections from many objects in the environment. The evaluation demonstrated the system ability to localise and image relatively large objects such as desktops, and couches, or objects with high reflective properties such as metallic surfaces. Smaller objects with low reflective properties have smaller cross-sections and thus reflect a smaller fraction of the Wi-Fi signals, which make them harder to image. Moreover, when the size of the object becomes close to the wavelength of the Wi-Fi signal, which is 12 cm approximately at 2.4 GHz, the interaction of the object with the Wi-Fi signals decreases. This is a fundamental limitation of imaging based on Wi-Fi signals. This fundamental limitation could be addressed using higher Wi-Fi frequencies such as 5 GHz that has a smaller wavelength of 6 cm approximately.

Sobron et al. [160, 161] presented a survey on different machine learning models used in human activity recognition and people counting. They also presented a new dataset for people counting applications with CSI measurements in several indoor environments such as rooms, corridors, and stairs. The dataset could be used to evaluate the performance of different machine learning models.

## 4. Discussion and future directions

During the course of this paper, some challenges facing Wi-Fi sensing systems and their applications have been identified. These systems still need to address some challenges in order to be able to operate in real-world environments, some of these challenges include, the presence of multipath propagation. The occlusion of the Wi-Fi signal. The presence of a large number of people. Furthermore, many of the proposed systems work well only in controlled environments; the accuracy of these systems is affected by many factors such as changes in distance and orientation of transceivers, human motions in surrounding areas, etc. Finally, Wi-Fi has limited range resolution in comparison with other sensing technology such as UWB, which could limit the range of applications.

The application of deep learning in sensing systems represents an interesting research direction that is still in its initial stage. Deep learning is a very appealing option because it has a remarkable ability in extracting useful features and then uses these features in different classification tasks. Although some recent works have shown promising results [162-167], some challenges still need more investigations. Further research must be conducted to propose deep learning architectures that best suits sensing systems.

## 5. Conclusions

This paper has presented a survey on different applications of the Wi-Fi based sensing systems such as elderly people monitoring, activity classification, gesture recognition, people counting, through the wall sensing, behind the corner sensing, and many other applications. The limitations of existing works were also highlighted along with many interesting future research directions.